\documentclass{article}

\usepackage{arxiv}

\usepackage[utf8]{inputenc} 
\usepackage[T1]{fontenc}    
\usepackage{hyperref}       
\usepackage{url}            
\usepackage{booktabs}       
\usepackage{amsfonts}       
\usepackage{amssymb}        
\usepackage{amsmath}        
\usepackage{nicefrac}       
\usepackage{microtype}      
\usepackage{lipsum}		
\usepackage{graphicx}
\usepackage[numbers]{natbib}
\usepackage{doi}
\usepackage{wrapfig}

\title{A systematic search for a three-velocity gyrodistributive law in special relativity with the {\tt lorentz} R package}

\author{ \href{https://orcid.org/0000-0001-5982-0415}{\includegraphics[width=0.03\textwidth]{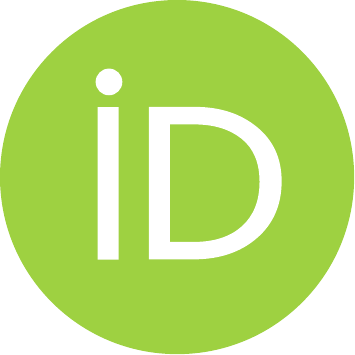}\hspace{1mm}Robin K. S.~Hankin}\thanks{\href{https://academics.aut.ac.nz/robin.hankin}{work};  
\href{https://www.youtube.com/watch?v=JzCX3FqDIOc&list=PL9_n3Tqzq9iWtgD8POJFdnVUCZ_zw6OiB&ab_channel=TrinTragulaGeneralRelativity}{play}} \\
 Auckland University of Technology\\
	\texttt{hankin.robin@gmail.com} \\
}

\renewcommand{\shorttitle}{Special relativity in R}

\hypersetup{
pdftitle={Special relativity in R},
pdfsubject={q-bio.NC, q-bio.QM},
pdfauthor={Robin K. S.~Hankin},
pdfkeywords={Special relativity, three-velocity, Thomas rotation,
  Lorentz transform, Poincare group, distributive law Einstein
  velocity addition, Wigner rotation, gyrogroup, gyromorphism,
  gyrocommutative, gyroassociative, four velocity, three-velocity,
  nonassociative, noncommutative}
}

\usepackage{Sweave}
\begin{document}
\maketitle

\begin{abstract}

Here I present the {\tt lorentz} package for working with relativistic
physics.  The package includes functionality for four-vector
transformations, three-velocity addition, and other relativistic
processes such as the behaviour of photons. It was designed to
facilitate the search for a gyrodistributive law.  In special
relativity, three-velocities and scalars constitute a gyrovector space
with addition $\oplus$ and scalar multiplication $\odot$.  Standard
vector spaces obey the distributive law $a(x+y)=ax+ay$ for scalar $a$
and vectors $x,y$; but no analogous gyrodistributive law for
$r\odot(u\oplus v)$ is known.  The package was designed to facilitate
the search for a gyrodistributive law and includes functionality for
four-vector transformations and three-velocity addition, which is
noncommutative and nonassociative.  I use the package to
systematically sweep a large space of potential gyrodistributive laws,
without success.  The package is available on CRAN, at
\url{https://CRAN.R-project.org/package=lorentz}.  \end{abstract}

\newcommand{\bu}{\mathbf u}
\newcommand{\bv}{\mathbf v}
\newcommand{\bw}{\mathbf w}
\newcommand{\bx}{\mathbf x}
\newcommand{\by}{\mathbf y}

\newcommand{\gyr}[2]{\operatorname{gyr}\left[{\mathbf #1},{\mathbf #2}\right]}

\setlength{\intextsep}{0pt}
\begin{wrapfigure}{r}{0.2\textwidth}
  \begin{center}
\includegraphics[width=1in]{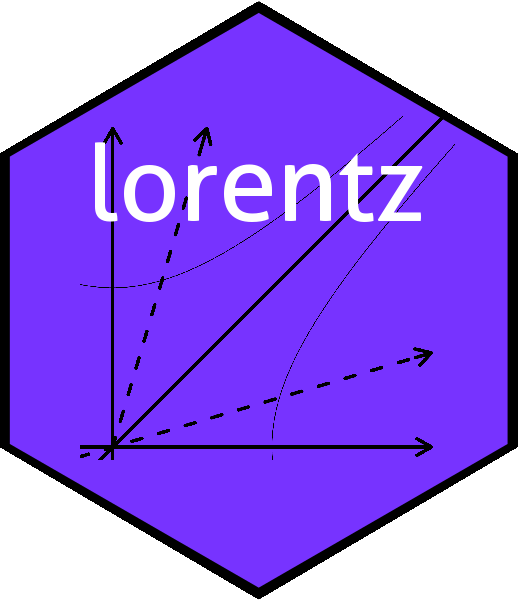}
  \end{center}
\end{wrapfigure}

\section{Introduction}

In special relativity, the Lorentz transforms supersede their
classical equivalent, the Galilean transforms~\citep{goldstein1980}.
Lorentz transforms operate on four-vectors such as the four-velocity
or four-potential and are usually operationalised as multiplication by
a $4\times 4$ matrix.  A Lorentz transform takes the components of an
arbitrary four-vector as observed in one coordinate system and returns
the components observed in another system which is moving at constant
velocity with respect to the first.

There are a few existing software tools for working with Lorentz
transforms, mostly developed in an educational context.  Early work
would include that of \citet{horwitz1992} who describe {\tt relLab}, a
system for building a range of {\em gendanken} experiments in an
interactive graphical environment.  The author asserts that it runs on
``any Macintosh computer with one megabyte of RAM or more'' but it is
not clear whether the software is still available.  More modern
contributions would include the {\tt OpenRelativity}
toolkit~\citep{sherin2016} which simulates the effects of special
relativity in the {\tt Unity} game engine.

The {\tt lorentz} package~\cite{hankin2022_lorentz_package} is written
in the R programming language~\cite{rcore2022}, providing {\tt
R}-centric functionality for the physics of special relativity.  It
deals with formal Lorentz boosts, converts between three-velocities
and four-velocities, and provides computational support for the
gyrogroup structure of relativistic three-velocity addition.  I
leverage the power of the R programming language and the package
itself to search for a gyrodistributive law in appendix A.

\section{The R programming language}

The R programming language~\cite{rcore2022} has an emphasis on
statistics and data analysis~\cite{chambers2008}.  However, R is a
general-purpose tool and is increasingly being used in the physical
sciences~\cite{mullen2022}.  For example, functionality for working
with general relativity is given
in~\cite{hankin2020,hankin2021string}.  R is interpreted, not
compiled, giving instant feedback on commands and allowing rapid
development.  In this document, the typical cycle is presented as
follows:

\begin{Schunk}
\begin{Sinput}
> 2+2
\end{Sinput}
\begin{Soutput}
[1] 4
\end{Soutput}
\end{Schunk}

We see the user's query of {\tt 2+2} is accepted, and the result, {\tt
4}, given as a response.  The ``{\tt [1]}'' indicates that the
returned value is a vector of length 1.

\subsection{The {\tt lorentz} package: an overview}

R's capabilities are extended through user-created ``packages'', which
offer specialist additional functionality to base R.  Packages may be
installed independently and cover a wide range of computational
facilities.  R packages contain code, data, and documentation in a
standardised format that can be installed by users of R, typically via
a centralised software repository such as CRAN.  R packages must
conform to a strict specification and pass extensive quality control
checks which ensure the usability and long-term stability of packages
for end users.

\subsection{Installation of the {\tt lorentz} package}

The R system may be downloaded from \url{https://cran.r-project.org/};
many users prefer the Rstudio IDE, available from
\url{https://posit.co/}.  Once R is installed, the {\tt lorentz}
package is easily loaded.  Type:

\begin{Schunk}
\begin{Sinput}
> install.packages("lorentz")
\end{Sinput}
\end{Schunk}

at the command line, and this will download the package from CRAN.  To
install the package, use

\begin{Schunk}
\begin{Sinput}
> library("lorentz")
\end{Sinput}
\end{Schunk}

and this will make the package functions available to the R session.

\section{Lorentz transforms: active and passive}

Passive transforms are the usual type of transforms taught and used in
relativity.  However, sometimes active transforms are needed and it is
easy to confuse the two.  Here I will discuss passive and then active
transforms, and illustrate both in a computational context.

\subsection*{Passive transforms}

\newcommand{\vvec}[2]{\begin{pmatrix}#1 \\ #2\end{pmatrix}}
\newcommand{\twomat}[4]{\begin{pmatrix} #1 & #2 \\ #3 & #4\end{pmatrix}}

Consider the following canonical Lorentz transform in which we have
motion in the $x$-direction at speed $v>0$; the motion is from left to
right.  We consider only the first two components of four-vectors, the
$y$- and $z$- components being trivial.  A typical physical
interpretation is that I am at rest, and my friend is in his spaceship
moving at speed $v$ past me; and we are wondering what vectors which I
measure in my own rest frame look like to him.  The (passive) Lorentz
transform is:

\begin{equation*}
\twomat{\gamma}{-\gamma v}{-\gamma v}{\gamma}
\end{equation*}

And the canonical example of that would be:

\begin{equation*}
\twomat{\gamma}{-\gamma v}{-\gamma v}{\gamma}\vvec{1}{0}=\vvec{\gamma}{-\gamma v}
\end{equation*}

where the vectors are four velocities (recall that $\vvec{1}{0}$ is
the four-velocity of an object at rest).  Operationally, I measure the
four-velocity of an object to be $\vvec{1}{0}$, and he measures the
same object as having a four-velocity of $\vvec{\gamma}{-\gamma v}$.
So I see the object at rest, and he sees it as moving at speed $-v$;
that is, he sees it moving to the left (it moves to the left because
he is moving to the right relative to me).  The {\tt lorentz}
package~\cite{hankin2022_lorentz_package} makes computations easy.
Suppose $v=0.6c$ in the $x$-direction.

\begin{Schunk}
\begin{Sinput}
>                           # NB: speed of light = 1 by default
> u <- as.3vel(c(0.6,0,0))  # coerce to a three-velocity
> u
\end{Sinput}
\begin{Soutput}
A vector of three-velocities (speed of light = 1)
       x y z
[1,] 0.6 0 0
\end{Soutput}
\begin{Sinput}
> as.4vel(u)                # four-velocity is better for calculations
\end{Sinput}
\begin{Soutput}
A vector of four-velocities (speed of light = 1)
        t    x y z
[1,] 1.25 0.75 0 0
\end{Soutput}
\begin{Sinput}
> (B <- boost(u))           # transformation matrix
\end{Sinput}
\begin{Soutput}
      t     x y z
t  1.25 -0.75 0 0
x -0.75  1.25 0 0
y  0.00  0.00 1 0
z  0.00  0.00 0 1
\end{Soutput}
\end{Schunk}

(note that element $[1,2]$ of the boost matrix $B$ is negative as we
have a passive transform).  Then a four-velocity of $(1,0,0,0)^T$
would appear in the moving frame as

\begin{Schunk}
\begin{Sinput}
> B %*% c(1,0,0,0)
\end{Sinput}
\begin{Soutput}
   [,1]
t  1.25
x -0.75
y  0.00
z  0.00
\end{Soutput}
\end{Schunk}

This corresponds to a speed of $-0.75/1.25=-0.6$.  Observe that it is
possible to transform an arbitrary four-vector:

\begin{Schunk}
\begin{Sinput}
> B %*% c(4,6,-8,9)
\end{Sinput}
\begin{Soutput}
  [,1]
t  0.5
x  4.5
y -8.0
z  9.0
\end{Soutput}
\end{Schunk}

\subsubsection*{Null vectors: light}

Let's try it with light (see section~\ref{photonsection} for more
details on photons).  Recall that we describe a photon in terms of its
four momentum, not four-velocity, which is undefined for a photon.
Specifically, we {\em define} the four-momentum of a photon to be

\begin{equation*}
  \left(
  \begin{array}{c}
    E/c\\Ev_x/c^2\\Ev_y/c^2\\Ev_z/c^2
  \end{array}
  \right)
  \end{equation*}

So if we consider unit energy and keep $c=1$ we get $p=\vvec{1}{1}$
in our one-dimensional world (for a rightward-moving photon) and the
Lorentz transform is then

\begin{equation*}
  \twomat{\gamma}{-\gamma v}{-\gamma v}{\gamma}\vvec{1}{1}=\vvec{\gamma-\gamma v}{\gamma-\gamma v}
  \end{equation*}

So, in the language used above, I see a photon with unit energy, and
my friend sees the photon with energy
$\gamma(1-v)=\sqrt{\frac{1-v}{1+v}}<1$, provided that $v>0$: the
photon has less energy in his frame than mine because of Doppler
redshifting.  It's worth doing the same analysis with a
leftward-moving photon:

\begin{equation*}
  \twomat{\gamma}{-\gamma v}{-\gamma v}{\gamma}\vvec{1}{-1}=\vvec{\gamma(1+v)}{-\gamma(1+v)}
  \end{equation*}

Here the photon has more energy for him than me because of blue
shifting: he is moving to the right and encounters a photon moving to
the left.  The R idiom would be

\begin{Schunk}
\begin{Sinput}
> B %*% c(1,1,0,0)
\end{Sinput}
\begin{Soutput}
  [,1]
t  0.5
x  0.5
y  0.0
z  0.0
\end{Soutput}
\begin{Sinput}
> B %*% c(1,-1,0,0)
\end{Sinput}
\begin{Soutput}
  [,1]
t    2
x   -2
y    0
z    0
\end{Soutput}
\end{Schunk}

for the left- and right- moving photons respectively. 

The above analysis uses {\em passive} transforms: there is a single
physical reality, and we describe that one physical reality using two
different coordinate systems.  One of the coordinate systems uses a
set of axes that are {\em boosted} relative to the axes of the other.

This is why it makes sense to use prime notation as in
$x\longrightarrow x'$ and $t\longrightarrow t'$ for a passive Lorentz
transform: the prime denotes measurements made using coordinates
that are defined with respect to the boosted system, and we see
notation like

\begin{equation*}
\vvec{t'}{x'}=\twomat{\gamma}{-\gamma v}{-\gamma v}{\gamma}\vvec{t}{x}
\end{equation*}  

These are the first two elements of a displacement four-vector.  It is
the same four-vector but viewed in two different reference frames.

\subsection*{Active transforms}

In the passive view, there is a single physical reality, and we are
just describing that one physical reality using two different
coordinate systems.  Now we will consider active transforms: there are
two physical realities, but one is boosted with respect to another.  

Suppose me and my friend have zero relative velocity, but my friend is
in a spaceship and I am outside it, in free space, at rest.  He
constructs a four-vector in his spaceship; for example, he could fire
bullets out of a gun which is fixed in the spaceship, and then
calculate their four-velocity as it appears to him in his
spaceship-centric coordinate system.  We both agree on this
four-velocity as our reference frames are identical: we have no
relative velocity.

Now his spaceship acquires a constant velocity, leaving me stationary.
My friend continues to fire bullets out of his gun and sees that their
four-velocity, as viewed in his spaceship coordinates, is the same as
when we were together.

Now he wonders what the four-velocity of the bullets is in my
reference frame.  This is an {\em active} transform: we have two
distinct physical realities, one in the spaceship when it was at rest
with respect to me, and one in the spaceship when moving.  And both
these realities, by construction, look the same to my friend in the
spaceship.

Suppose, for example, he sees the bullets at rest in his spaceship;
they have a four-velocity of $\vvec{1}{0}$, and my friend says to
himself: ``I see bullets with a four velocity of $\vvec{1}{0}$, and I
know what that means.  The bullets are at rest.  What are the bullets'
four velocities in Robin's reference frame?".  This is an {\em active}
transform:

\begin{equation*}
\twomat{\gamma}{\gamma v}{\gamma v}{\gamma}\vvec{1}{0}=\vvec{\gamma}{\gamma v}
\end{equation*}

(we again suppose that the spaceship moves at speed $v>0$ from left to
right).  So he sees a four velocity of $\vvec{1}{0}$ and I see
$\vvec{\gamma}{\gamma v}$, that is, with a positive speed: the bullets
move from left to right (with the spaceship).  The R idiom would be:

\begin{Schunk}
\begin{Sinput}
> (B <- boost(as.3vel(c(0.8,0,0))))  # 0.8c left to right
\end{Sinput}
\begin{Soutput}
          t         x y z
t  1.666667 -1.333333 0 0
x -1.333333  1.666667 0 0
y  0.000000  0.000000 1 0
z  0.000000  0.000000 0 1
\end{Soutput}
\begin{Sinput}
> solve(B) %*% c(1,0,0,0)            # active transform
\end{Sinput}
\begin{Soutput}
      [,1]
t 1.666667
x 1.333333
y 0.000000
z 0.000000
\end{Soutput}
\end{Schunk}

\section{Successive Lorentz transforms}

Coordinate transformation is effected by standard matrix
 multiplication; thus composition of two Lorentz transforms is also
 ordinary matrix multiplication:

\begin{Schunk}
\begin{Sinput}
> u <- as.3vel(c(0.3,-0.4,+0.8))
> v <- as.3vel(c(0.4,+0.2,-0.1))
> L <- boost(u) %*% boost(v)
> L
\end{Sinput}
\begin{Soutput}
          t          x          y          z
t  3.256577 -2.2327055  0.5419596 -2.0800479
x -1.437147  1.6996791 -0.0237489  0.4194255
y  1.091131 -0.7581795  1.1190282 -0.6029155
z -2.519789  1.5878378 -0.2023170  2.1879612
\end{Soutput}
\end{Schunk}

But observe that the resulting transform is not a pure boost, as the
spatial components are not symmetrical.  We may decompose the matrix
product $L$ into a pure translation composed with an orthogonal
matrix, which represents a coordinate rotation.  The R idiom is
{\tt pureboost()} for the pure boost component, and {\tt orthog()}
for the rotation:
             
\begin{Schunk}
\begin{Sinput}
> (P <- pureboost(L))  # pure boost
\end{Sinput}
\begin{Soutput}
           t          x          y          z
t  3.2565770 -2.2327055  0.5419596 -2.0800479
x -2.2327055  2.1711227 -0.2842745  1.0910491
y  0.5419596 -0.2842745  1.0690039 -0.2648377
z -2.0800479  1.0910491 -0.2648377  2.0164504
\end{Soutput}
\begin{Sinput}
> P - t(P)   # check for symmetry
\end{Sinput}
\begin{Soutput}
  t x y z
t 0 0 0 0
x 0 0 0 0
y 0 0 0 0
z 0 0 0 0
\end{Soutput}
\end{Schunk}

Now we compute the rotation:
\begin{Schunk}
\begin{Sinput}
> (U <- orthog(L))                  # rotation matrix
\end{Sinput}
\begin{Soutput}
              t             x            y             z
t  1.000000e+00 -1.332268e-14 3.219647e-15 -1.509903e-14
x -1.054712e-14  9.458514e-01 1.592328e-01 -2.828604e-01
y  8.659740e-15 -1.858476e-01 9.801022e-01 -6.971587e-02
z -1.953993e-14  2.661311e-01 1.185098e-01  9.566241e-01
\end{Soutput}
\begin{Sinput}
> U[2:4,2:4]                        # inspect the spatial components
\end{Sinput}
\begin{Soutput}
           x         y           z
x  0.9458514 0.1592328 -0.28286043
y -0.1858476 0.9801022 -0.06971587
z  0.2661311 0.1185098  0.95662410
\end{Soutput}
\begin{Sinput}
> round(crossprod(U) - diag(4),10)  # check for orthogonality
\end{Sinput}
\begin{Soutput}
  t x y z
t 0 0 0 0
x 0 0 0 0
y 0 0 0 0
z 0 0 0 0
\end{Soutput}
\begin{Sinput}
> ## zero to within numerical uncertainty
\end{Sinput}
\end{Schunk}

\section[Units in which c is not 1]{Units in which ${\mathbf c\neq 1}$}  

The preceding material used units in which $c=1$.  Here I show how the
package deals with units such as SI in which $c=299792458\neq 1$.  For
obvious reasons we cannot have a function called {\tt c()} so the
package gets and sets the speed of light with function {\tt sol()}:

\begin{Schunk}
\begin{Sinput}
> sol(299792458)
\end{Sinput}
\begin{Soutput}
[1] 299792458
\end{Soutput}
\begin{Sinput}
> sol()
\end{Sinput}
\begin{Soutput}
[1] 299792458
\end{Soutput}
\end{Schunk}

The speed of light is now~$299792458$ until re-set by {\tt sol()} (an
empty argument queries the speed of light).  We now consider speeds
which are fast by terrestrial standards but involve only a small
relativistic correction to the Galilean result:
                                                
\begin{Schunk}
\begin{Sinput}
> u <- as.3vel(c(100,200,300))
> as.4vel(u)
\end{Sinput}
\begin{Soutput}
A vector of four-velocities (speed of light = 299792458)
     t   x   y   z
[1,] 1 100 200 300
\end{Soutput}
\end{Schunk}

The gamma correction term $\gamma$ is only very slightly larger
than~$1$ and indeed R's default print method suppresses the
difference:
                    
\begin{Schunk}
\begin{Sinput}
> gam(u)
\end{Sinput}
\begin{Soutput}
[1] 1
\end{Soutput}
\end{Schunk}

However, we can display more significant figures by subtracting one:

\begin{Schunk}
\begin{Sinput}
> gam(u)-1
\end{Sinput}
\begin{Soutput}
[1] 7.789325e-13
\end{Soutput}
\end{Schunk}

or alternatively we can use the {\tt gamm1()} function which
calculates $\gamma-1$ more accurately for speeds $\ll c$:
                                                  
\begin{Schunk}
\begin{Sinput}
> gamm1(u)
\end{Sinput}
\begin{Soutput}
[1] 7.78855e-13
\end{Soutput}
\end{Schunk}

The Lorentz boost is again calculated by the {\tt boost()} function:
              
\begin{Schunk}
\begin{Sinput}
> boost(u)
\end{Sinput}
\begin{Soutput}
     t             x            y             z
t    1 -1.112650e-15 -2.22530e-15 -3.337950e-15
x -100  1.000000e+00  1.11265e-13  1.668975e-13
y -200  1.112650e-13  1.00000e+00  3.337950e-13
z -300  1.668975e-13  3.33795e-13  1.000000e+00
\end{Soutput}
\end{Schunk}

The boost matrix is not symmetrical, even though it is a pure boost,
because $c\neq 1$.

Note how the transform is essentially the Galilean
result, which is discussed below.

\subsection{Changing units}

Often we have a four-vector in SI units and wish to express this in natural units.

\begin{Schunk}
\begin{Sinput}
> sol(299792458)
\end{Sinput}
\begin{Soutput}
[1] 299792458
\end{Soutput}
\begin{Sinput}
> disp  <- c(1,1,0,0)
\end{Sinput}
\end{Schunk}

If we interpret {\tt disp} as a four-displacement, it corresponds to
moving 1 metre along the x-axis and waiting for one second.  To
convert this to natural units we multiply by the passive
transformation matrix given by {\tt ptm()}:

\begin{Schunk}
\begin{Sinput}
> ptm(to_natural=TRUE) %*% disp
\end{Sinput}
\begin{Soutput}
       [,1]
t 299792458
x         1
y         0
z         0
\end{Soutput}
\end{Schunk}

In the above, see how the same vector is expressed in natural units in
which the speed of light is equal to 1: the unit of time is about
$3\times 10^{-9}$ seconds and the unit of distance remains the metre.
Alternatively, we might decide to keep the unit of time equal to one
second, and use a unit of distance equal to 299792458 metres which
again ensures that~$c=1$:

\begin{Schunk}
\begin{Sinput}
> ptm(to_natural=TRUE,change_time=FALSE) %*% disp
\end{Sinput}
\begin{Soutput}
          [,1]
t 1.000000e+00
x 3.335641e-09
y 0.000000e+00
z 0.000000e+00
\end{Soutput}
\end{Schunk}

As a further check, we can take two boost matrices corresponding to
the same coordinate transformation but expressed using different units
of length and verify that their orthogonal component agrees:

\begin{Schunk}
\begin{Sinput}
> sol(1)
\end{Sinput}
\begin{Soutput}
[1] 1
\end{Soutput}
\begin{Sinput}
> B1 <- boost((2:4)/10) %*% boost(c(-5,1,3)/10)
> orthog(B1)[2:4,2:4]
\end{Sinput}
\begin{Soutput}
           x           y          z
x  0.9832336  0.09752166 0.15408208
y -0.1020390  0.99454439 0.02166761
z -0.1511284 -0.03702671 0.98782044
\end{Soutput}
\end{Schunk}

Now we create {\tt B2} which is the same physical object but using a
length scale of one-tenth of {\tt B2} (which requires that we
multiply the speed of light by a factor of 10):

\begin{Schunk}
\begin{Sinput}
> sol(10)
\end{Sinput}
\begin{Soutput}
[1] 10
\end{Soutput}
\begin{Sinput}
> B2 <- boost(2:4) %*% boost(c(-5,1,3))  # exactly the same as B1 above
> orthog(B2)[2:4,2:4]
\end{Sinput}
\begin{Soutput}
           x           y          z
x  0.9832336  0.09752166 0.15408208
y -0.1020390  0.99454439 0.02166761
z -0.1511284 -0.03702671 0.98782044
\end{Soutput}
\end{Schunk}

so the two matrices agree, as expected.

\section{Infinite speed of light}

In the previous section considered speeds that were small compared
with the speed of light and here we will consider the classical limit
of infinite $c$:

\begin{Schunk}
\begin{Sinput}
> sol(Inf)
\end{Sinput}
\begin{Soutput}
[1] Inf
\end{Soutput}
\end{Schunk}

Then the familiar parallelogram law operates:

\begin{Schunk}
\begin{Sinput}
> u <- as.3vel(1:3)
> v <- as.3vel(c(-6,8,3))
> u+v
\end{Sinput}
\begin{Soutput}
A vector of three-velocities (speed of light = Inf)
      x  y z
[1,] -5 10 6
\end{Soutput}
\begin{Sinput}
> v+u
\end{Sinput}
\begin{Soutput}
A vector of three-velocities (speed of light = Inf)
      x  y z
[1,] -5 10 6
\end{Soutput}
\end{Schunk}

Above we see that composition of velocities is commutative, unlike the
relativistic case.  The boost matrix is instructive:

\begin{Schunk}
\begin{Sinput}
> boost(u)
\end{Sinput}
\begin{Soutput}
   t x y z
t  1 0 0 0
x -1 1 0 0
y -2 0 1 0
z -3 0 0 1
\end{Soutput}
\begin{Sinput}
> boost(u+v)
\end{Sinput}
\begin{Soutput}
    t x y z
t   1 0 0 0
x   5 1 0 0
y -10 0 1 0
z  -6 0 0 1
\end{Soutput}
\begin{Sinput}
> boost(u) %*% boost(v)
\end{Sinput}
\begin{Soutput}
    t x y z
t   1 0 0 0
x   5 1 0 0
y -10 0 1 0
z  -6 0 0 1
\end{Soutput}
\end{Schunk}

Above, see how the boost matrix for the composed velocity of $u+v$
does not have any rotational component, unlike the relativistic case
[recall that {\tt boost()} gives a {\em passive} transform, which is
why the sign of the numbers in the first column is changed].  With an
infinite speed of light, even ``large'' speeds have zero relativistic
correction:

\begin{Schunk}
\begin{Sinput}
> gamm1(1e100)
\end{Sinput}
\begin{Soutput}
[1] 0
\end{Soutput}
\end{Schunk}

Function {\tt rboost()} returns a random Lorentz transform matrix,
which is in general a combination of a pure Lorentz boost and an
orthogonal rotation.  With an infinite speed of light, it requires a
speed:

\begin{Schunk}
\begin{Sinput}
> set.seed(0)
> options(digits=3)
> (B <- rboost(1))  # random boost, speed 1
\end{Sinput}
\begin{Soutput}
          t      x       y      z
[1,]  1.000  0.000  0.0000  0.000
[2,] -0.411  0.213 -0.9402  0.266
[3,] -0.279 -0.917 -0.0989  0.385
[4,]  0.868 -0.336 -0.3260 -0.884
\end{Soutput}
\end{Schunk}

We can decompose {\tt B} into a pure boost and an orthogonal
transformation:

\begin{Schunk}
\begin{Sinput}
> orthog(B)
\end{Sinput}
\begin{Soutput}
     [,1]   [,2]    [,3]   [,4]
[1,]    1  0.000  0.0000  0.000
[2,]    0  0.213 -0.9402  0.266
[3,]    0 -0.917 -0.0989  0.385
[4,]    0 -0.336 -0.3260 -0.884
\end{Soutput}
\begin{Sinput}
> pureboost(B)
\end{Sinput}
\begin{Soutput}
          t x y z
[1,]  1.000 0 0 0
[2,] -0.123 1 0 0
[3,]  0.131 0 1 0
[4,] -0.984 0 0 1
\end{Soutput}
\end{Schunk}

Boost matrices can be applied to any four-vector.  Here I show how
pure spatial displacements transform with an infinite light speed.

\begin{Schunk}
\begin{Sinput}
> (u <- as.3vel(c(10,0,0))) # velocity of 10, parallel to x axis
\end{Sinput}
\begin{Soutput}
A vector of three-velocities (speed of light = Inf)
      x y z
[1,] 10 0 0
\end{Soutput}
\begin{Sinput}
> (B <- boost(u))
\end{Sinput}
\begin{Soutput}
    t x y z
t   1 0 0 0
x -10 1 0 0
y   0 0 1 0
z   0 0 0 1
\end{Soutput}
\begin{Sinput}
> d <- c(0,1,0,0) # displacement of distance one, parallel to the x-axis
> B %*% d
\end{Sinput}
\begin{Soutput}
  [,1]
t    0
x    1
y    0
z    0
\end{Soutput}
\end{Schunk}

Above we see that a spatial displacement is the same for both
observers.  We can similarly apply a boost to a temporal displacement:

\begin{Schunk}
\begin{Sinput}
> d <- c(1,0,0,0) # displacement of one unit of time, no spatial component
> B %*% d
\end{Sinput}
\begin{Soutput}
  [,1]
t    1
x  -10
y    0
z    0
\end{Soutput}
\end{Schunk}

Above we see the result expected from classical mechanics.

\section{Vectorization}

Here I discuss vectorized operations (to avoid confusion between boost
matrices and their transposes we will use $c=10$).  The issue is
difficult because a Lorentz boost is conceptually a matrix product of
a $4\times 4$ matrix with vector with four elements:

\begin{Schunk}
\begin{Sinput}
> sol(10)
\end{Sinput}
\begin{Soutput}
[1] 10
\end{Soutput}
\begin{Sinput}
> u <- as.3vel(c(5,-6,4))
> (U <- as.4vel(u))
\end{Sinput}
\begin{Soutput}
A vector of four-velocities (speed of light = 10)
        t    x     y    z
[1,] 2.09 10.4 -12.5 8.34
\end{Soutput}
\begin{Sinput}
> B <- boost(U)
> B %*% as.vector(U)
\end{Sinput}
\begin{Soutput}
       [,1]
t  1.00e+00
x -1.33e-15
y  4.44e-16
z  1.78e-15
\end{Soutput}
\end{Schunk}

(note that the result is the four-velocity of an object at rest, as
expected, for we use passive transforms by default).  However, things
are different if we wish to consider many four-vectors in one R
object.  A vector ${\mathbf V}$ of four-velocities is a matrix: each
{\em row} of ${\mathbf V}$ is a four-velocity.  In the package we
represent this with objects of class {\tt 4vel}.  Because a vector is
treated (almost) as a one-column matrix in R, and the four-velocities
are rows, we need to take a transpose in some sense.

\begin{Schunk}
\begin{Sinput}
> u <- 1:7  # speed in the x-direction [c=10]
> jj <- cbind(gam(u),gam(u)*u,0,0)
> (U <- as.4vel(jj))
\end{Sinput}
\begin{Soutput}
A vector of four-velocities (speed of light = 10)
        t    x y z
[1,] 1.01 1.01 0 0
[2,] 1.02 2.04 0 0
[3,] 1.05 3.14 0 0
[4,] 1.09 4.36 0 0
[5,] 1.15 5.77 0 0
[6,] 1.25 7.50 0 0
[7,] 1.40 9.80 0 0
\end{Soutput}
\end{Schunk}

Now a boost, also in the x-direction:

\begin{Schunk}
\begin{Sinput}
> (B <- boost(as.3vel(c(6,0,0))))  # 60% speed of light
\end{Sinput}
\begin{Soutput}
      t      x y z
t  1.25 -0.075 0 0
x -7.50  1.250 0 0
y  0.00  0.000 1 0
z  0.00  0.000 0 1
\end{Soutput}
\end{Schunk}

Note the asymmetry of $B$, in this case reflecting the speed of light
being 10 (but note that boost matrices are not always symmetrical,
even if $c=1$).

To effect a {\em passive} boost we need to multiply each row of $U$ by
the transpose of the boost matrix $B$:

\begin{Schunk}
\begin{Sinput}
> U %*% t(B)
\end{Sinput}
\begin{Soutput}
        t     x y z
[1,] 1.18 -6.28 0 0
[2,] 1.12 -5.10 0 0
[3,] 1.07 -3.93 0 0
[4,] 1.04 -2.73 0 0
[5,] 1.01 -1.44 0 0
[6,] 1.00  0.00 0 0
[7,] 1.02  1.75 0 0
\end{Soutput}
\end{Schunk}

we can verify that the above is at least plausible:

\begin{Schunk}
\begin{Sinput}
> is.consistent.4vel(U %*% t(B))
\end{Sinput}
\begin{Soutput}
[1] TRUE TRUE TRUE TRUE TRUE TRUE TRUE
\end{Soutput}
\end{Schunk}

the above shows that the four velocities $U$, as observed by an
observer corresponding to boost $B$, satisfies $U^iU_i=-c^2$.  Anyway,
in this context we really ought to use {\tt tcrossprod()}:

\begin{Schunk}
\begin{Sinput}
> tcrossprod(U,B)
\end{Sinput}
\begin{Soutput}
        t     x y z
[1,] 1.18 -6.28 0 0
[2,] 1.12 -5.10 0 0
[3,] 1.07 -3.93 0 0
[4,] 1.04 -2.73 0 0
[5,] 1.01 -1.44 0 0
[6,] 1.00  0.00 0 0
[7,] 1.02  1.75 0 0
\end{Soutput}
\end{Schunk}

which would be preferable (because this idiom does not require one to
take a transpose) although the speed increase is unlikely to matter
much because $B$ is only $4\times 4$.

The above transforms were passive: we have some four-vectors measured
in my rest frame, and we want to see what these are four-vectors as
measured by my friend, who is moving in the positive x direction at
60\% of the speed of light (remember that $c=10$).  See how the
x-component of the transformed four-velocity is negative, because in
my friend's rest frame, the four velocities are pointing backwards.

To effect an {\em active} transform we need to take the matrix inverse
of $B$:

\begin{Schunk}
\begin{Sinput}
> solve(B)
\end{Sinput}
\begin{Soutput}
     t     x y z
t 1.25 0.075 0 0
x 7.50 1.250 0 0
y 0.00 0.000 1 0
z 0.00 0.000 0 1
\end{Soutput}
\end{Schunk}

and then

\begin{Schunk}
\begin{Sinput}
> tcrossprod(U,solve(B))
\end{Sinput}
\begin{Soutput}
        t     x y z
[1,] 1.33  8.79 0 0
[2,] 1.43 10.21 0 0
[3,] 1.55 11.79 0 0
[4,] 1.69 13.64 0 0
[5,] 1.88 15.88 0 0
[6,] 2.13 18.75 0 0
[7,] 2.49 22.75 0 0
\end{Soutput}
\end{Schunk}

In the above, note how the positive x-component of the four-velocity
is increased because we have actively boosted it.  We had better check
the result for consistency:

\begin{Schunk}
\begin{Sinput}
> is.consistent.4vel(tcrossprod(U,solve(B)))
\end{Sinput}
\begin{Soutput}
[1] TRUE TRUE TRUE TRUE TRUE TRUE TRUE
\end{Soutput}
\end{Schunk}

\section{Multiple boosts}

If we are considering multiple boosts, it is important to put them in
the correct order.  First we will do some passive boosts.

\begin{Schunk}
\begin{Sinput}
> sol(100)
\end{Sinput}
\begin{Soutput}
[1] 100
\end{Soutput}
\begin{Sinput}
> B1 <- boost(r3vel(1)) %*% boost(r3vel(1))
> B2 <- boost(r3vel(1)) %*% boost(r3vel(1)) 
> (U <- r4vel(5))
\end{Sinput}
\begin{Soutput}
A vector of four-velocities (speed of light = 100)
        t      x     y      z
[1,] 1.99 -162.4  46.3  34.32
[2,] 2.58  149.0 185.0  -3.07
[3,] 1.75 -110.4 -18.4 -90.19
[4,] 1.70   55.1 -88.3  90.15
[5,] 2.62 -205.6  98.2  81.53
\end{Soutput}
\end{Schunk}

Successive boosts are effected by matrix multiplication; there are at
least four equivalent R constructions:

\begin{Schunk}
\begin{Sinput}
> U %*% t(B1) %*% t(B2)
\end{Sinput}
\begin{Soutput}
         t    x     y    z
[1,] 11.09 -601  -844 -383
[2,]  3.00  -70  -166 -218
[3,] 13.98 -639 -1087 -594
[4,]  7.75 -239  -697 -220
[5,] 12.23 -706  -911 -394
\end{Soutput}
\begin{Sinput}
> U %*% t(B2 %*% B1)    # note order of operations
\end{Sinput}
\begin{Soutput}
         t    x     y    z
[1,] 11.09 -601  -844 -383
[2,]  3.00  -70  -166 -218
[3,] 13.98 -639 -1087 -594
[4,]  7.75 -239  -697 -220
[5,] 12.23 -706  -911 -394
\end{Soutput}
\begin{Sinput}
> tcrossprod(U, B2 %*% B1)
\end{Sinput}
\begin{Soutput}
         t    x     y    z
[1,] 11.09 -601  -844 -383
[2,]  3.00  -70  -166 -218
[3,] 13.98 -639 -1087 -594
[4,]  7.75 -239  -697 -220
[5,] 12.23 -706  -911 -394
\end{Soutput}
\begin{Sinput}
> U %>% tcrossprod(B2 %*% B1)
\end{Sinput}
\begin{Soutput}
         t    x     y    z
[1,] 11.09 -601  -844 -383
[2,]  3.00  -70  -166 -218
[3,] 13.98 -639 -1087 -594
[4,]  7.75 -239  -697 -220
[5,] 12.23 -706  -911 -394
\end{Soutput}
\end{Schunk}

(in the above, note that the result is the same in each case).

\subsection*{A warning}

It is easy to misapply matrix multiplication in this context.  Note
carefully that the following natural idiom is {\bf incorrect}:

\begin{Schunk}
\begin{Sinput}
> U %*% B  # Young Frankstein: Do Not Use This Brain!
\end{Sinput}
\begin{Soutput}
         t      x     y      z
[1,]  1220 -203.2  46.3  34.32
[2,] -1115  186.1 185.0  -3.07
[3,]   830 -138.1 -18.4 -90.19
[4,]  -411   68.7 -88.3  90.15
[5,]  1545 -257.1  98.2  81.53
\end{Soutput}
\begin{Sinput}
> ## The above idiom is incorrect.  See
> ## https://www.youtube.com/watch?v=m7-bMBuVmHo&t=1s
> ## (in particular @1:08) for a technical explanation of why 
> ## this is a Very Bad Idea (tm).
\end{Sinput}
\end{Schunk}

It is not clear to me that the idiom above has any meaning at all.

\section{The stress-energy tensor}

The stress-energy tensor (sometimes the energy-momentum tensor) is a
generalization and combination of the classical concepts of density,
energy flux, and the classical stress tensor~\citep{schutz1985}.  It
is a contravariant tensor of rank two, usually represented as a
symmetric $4\times 4$ matrix.  The {\tt lorentz} package includes
functionality for applying Lorentz transforms to the stress energy
tensor.

\begin{Schunk}
\begin{Sinput}
> sol(1)        # revert to natural units 
\end{Sinput}
\begin{Soutput}
[1] 1
\end{Soutput}
\begin{Sinput}
> D <- dust(1)  # Dust is the simplest nontrivial SET, with 
> D             # only one nonzero component
\end{Sinput}
\begin{Soutput}
  t x y z
t 1 0 0 0
x 0 0 0 0
y 0 0 0 0
z 0 0 0 0
\end{Soutput}
\end{Schunk}

The stress-energy tensor is usually written with two upstairs
(contravariant) indices, as in~$T^{\alpha\beta}$; it may be
transformed using the {\tt transform\_uu()} function: package:

\begin{Schunk}
\begin{Sinput}
> B <- boost(as.3vel(c(0.0,0.8,0.0)))
> transform_uu(D,B)
\end{Sinput}
\begin{Soutput}
      t x     y z
t  2.78 0 -2.22 0
x  0.00 0  0.00 0
y -2.22 0  1.78 0
z  0.00 0  0.00 0
\end{Soutput}
\end{Schunk}

In this reference frame, the dust is not at rest: the stress-energy
tensor has components corresponding to nonzero pressure and momentum
transfer, and the $[t,t]$ component is greater, at 2.78, than its rest
value of 1.  Note that the $[t,y]$ component is negative as we use
passive transforms.  If one wants to consider the stress-energy tensor
with downstairs indices (here we will use a photon gas), we need to
use {\tt transform\_dd()}:

\begin{Schunk}
\begin{Sinput}
> pg <- photongas(3)
> pg
\end{Sinput}
\begin{Soutput}
  t x y z
t 3 0 0 0
x 0 1 0 0
y 0 0 1 0
z 0 0 0 1
\end{Soutput}
\begin{Sinput}
> transform_uu(pg,B)
\end{Sinput}
\begin{Soutput}
      t x     y z
t 10.11 0 -8.89 0
x  0.00 1  0.00 0
y -8.89 0  8.11 0
z  0.00 0  0.00 1
\end{Soutput}
\end{Schunk}

again we see that the $[0,0]$ component is larger than its rest value,
and we see nonzero off-diagonal components which correspond to the
dynamical behaviour.  As a consistency check we can verify that this
is the same as transforming the SET with upstairs indices, using the
{\tt lower()} and {\tt raise()} functions:

\begin{Schunk}
\begin{Sinput}
> raise(transform_dd(lower(pg),lower(B)))
\end{Sinput}
\begin{Soutput}
      t x     y z
t 10.11 0 -8.89 0
x  0.00 1  0.00 0
y -8.89 0  8.11 0
z  0.00 0  0.00 1
\end{Soutput}
\begin{Sinput}
> raise(transform_dd(lower(pg),lower(B))) - transform_uu(pg,B) #zero to numerical precision
\end{Sinput}
\begin{Soutput}
  t x y z
t 0 0 0 0
x 0 0 0 0
y 0 0 0 0
z 0 0 0 0
\end{Soutput}
\end{Schunk}

One of the calls to {\tt lower()} is redundant; for a photon gas,
raising or lowering both indices does not change the components as the
Minkowski metric is symmetric and orthogonal.

\subsection{Successive boosts}
  
Successive boots are represented as ordinary matrix multiplication.
Again the {\tt magrittr} package can be used for more readable idiom.
 
\begin{Schunk}
\begin{Sinput}
> B1 <- boost(as.3vel(c(0.5,-0.4,0.6)))
> B2 <- boost(as.3vel(c(0.1,-0.1,0.3)))
> pf <- perfectfluid(4,1)
> pf
\end{Sinput}
\begin{Soutput}
  t x y z
t 4 0 0 0
x 0 1 0 0
y 0 0 1 0
z 0 0 0 1
\end{Soutput}
\begin{Sinput}
> pf %>% transform_uu(B1) %>% transform_uu(B2)
\end{Sinput}
\begin{Soutput}
      t      x      y     z
t  38.4 -18.17  15.24 -28.2
x -18.2   9.38  -7.03  13.0
y  15.2  -7.03   6.89 -10.9
z -28.2  12.98 -10.89  21.1
\end{Soutput}
\begin{Sinput}
> pf %>% transform_uu(B2 %*% B1)  # should match
\end{Sinput}
\begin{Soutput}
      t      x      y     z
t  38.4 -18.17  15.24 -28.2
x -18.2   9.38  -7.03  13.0
y  15.2  -7.03   6.89 -10.9
z -28.2  12.98 -10.89  21.1
\end{Soutput}
\end{Schunk}

Again as a consistency check, we may verify that transforming
downstairs indices gives the same result:

\begin{Schunk}
\begin{Sinput}
> lower(pf) %>% transform_dd(lower(B1) %*% lower(B2)) %>% raise()
\end{Sinput}
\begin{Soutput}
      t      x      y     z
t  38.4 -18.17  15.24 -28.2
x -18.2   9.38  -7.03  13.0
y  15.2  -7.03   6.89 -10.9
z -28.2  12.98 -10.89  21.1
\end{Soutput}
\end{Schunk}

(note that the matrix representation of the Lorentz transforms
requires that the order of multiplication be reversed for successive
covariant transforms, so {\tt B1} and {\tt B2} must be swapped).

\subsection{Speed of light and the stress-energy tensor}

Here I will perform another consistency check, this time with non-unit
speed of light, for a perfect fluid:

\begin{Schunk}
\begin{Sinput}
> sol(10)
\end{Sinput}
\begin{Soutput}
[1] 10
\end{Soutput}
\begin{Sinput}
> pf_rest <- perfectfluid(1,4)
> pf_rest
\end{Sinput}
\begin{Soutput}
     t    x    y    z
t 1.04 0.00 0.00 0.00
x 0.00 0.04 0.00 0.00
y 0.00 0.00 0.04 0.00
z 0.00 0.00 0.00 0.04
\end{Soutput}
\end{Schunk}

Thus {\tt pf\_rest} is the stress energy for a perfect fluid at rest
in a particular frame $F$.  We may now consider the same perfect
fluid, but moving with a three velocity of~$(3,4,5)'$: with respect to
$F$:

\begin{Schunk}
\begin{Sinput}
> u <- as.3vel(3:5)
> pf_moving <- perfectfluid(1,4,u)
> pf_moving
\end{Sinput}
\begin{Soutput}
      t     x     y    z
t  2.08  6.24  8.32 10.4
x  6.24 18.76 24.96 31.2
y  8.32 24.96 33.32 41.6
z 10.40 31.20 41.60 52.0
\end{Soutput}
\end{Schunk}

The consistency check is to verify that transforming to a frame in
which the fluid is at rest will result in a stress-energy tensor that
matches {\tt pf\_rest}:

\begin{Schunk}
\begin{Sinput}
> transform_uu(perfectfluid(1,4,u),boost(u))
\end{Sinput}
\begin{Soutput}
          t         x         y         z
t  1.04e+00 -3.01e-16 -1.65e-15  9.04e-16
x -1.33e-15  4.00e-02 -1.87e-15 -4.95e-15
y -3.33e-15  7.18e-15  4.00e-02 -1.87e-15
z -3.55e-15  9.87e-16  1.08e-14  4.00e-02
\end{Soutput}
\end{Schunk}

thus showing agreement to within numerical precision.

\section{Photons}
\label{photonsection}
It is possible to define the four-momentum of photons by specifying
their three-velocity and energy, and using {\tt as.photon()}:
 
\begin{Schunk}
\begin{Sinput}
> sol(1)
\end{Sinput}
\begin{Soutput}
[1] 1
\end{Soutput}
\begin{Sinput}
> (A <- as.photon(as.3vel(cbind(0.9,1:5/40,5:1/40))))
\end{Sinput}
\begin{Soutput}
     E   p_x    p_y    p_z
[1,] 1 0.990 0.0275 0.1375
[2,] 1 0.992 0.0551 0.1103
[3,] 1 0.993 0.0828 0.0828
[4,] 1 0.992 0.1103 0.0551
[5,] 1 0.990 0.1375 0.0275
\end{Soutput}
\end{Schunk}

above, $A$ is a vector of four-momentum of five photons, all of unit
energy, each with a null world line.  They are all moving
approximately parallel to the x-axis.  We can check that this is
indeed a null vector:
                                                 
\begin{Schunk}
\begin{Sinput}
> inner4(A)
\end{Sinput}
\begin{Soutput}
[1]  1.45e-16  2.56e-17 -5.55e-17  2.56e-17  1.45e-16
\end{Soutput}
\end{Schunk}

showing that the vectors are indeed null to numerical precision.  What
do these photons look like in a frame moving along the $x$-axis at
$0.7c$?
  
\begin{Schunk}
\begin{Sinput}
> tcrossprod(A,boost(as.3vel(c(0.7,0,0))))
\end{Sinput}
\begin{Soutput}
         t     x      y      z
[1,] 0.430 0.406 0.0275 0.1375
[2,] 0.428 0.409 0.0551 0.1103
[3,] 0.427 0.410 0.0828 0.0828
[4,] 0.428 0.409 0.1103 0.0551
[5,] 0.430 0.406 0.1375 0.0275
\end{Soutput}
\end{Schunk}

Above, see how the photons have lost the majority of their energy due
to redshifting.  Blue shifting is easy to implement as either a
passive transform:
                                            
\begin{Schunk}
\begin{Sinput}
> tcrossprod(A,boost(as.3vel(c(-0.7,0,0))))
\end{Sinput}
\begin{Soutput}
        t    x      y      z
[1,] 2.37 2.37 0.0275 0.1375
[2,] 2.37 2.37 0.0551 0.1103
[3,] 2.37 2.37 0.0828 0.0828
[4,] 2.37 2.37 0.1103 0.0551
[5,] 2.37 2.37 0.1375 0.0275
\end{Soutput}
\end{Schunk}
                  
or an active transform:

\begin{Schunk}
\begin{Sinput}
> tcrossprod(A,solve(boost(as.3vel(c(0.7,0,0)))))
\end{Sinput}
\begin{Soutput}
        t    x      y      z
[1,] 2.37 2.37 0.0275 0.1375
[2,] 2.37 2.37 0.0551 0.1103
[3,] 2.37 2.37 0.0828 0.0828
[4,] 2.37 2.37 0.1103 0.0551
[5,] 2.37 2.37 0.1375 0.0275
\end{Soutput}
\end{Schunk}

giving the same result.

\subsection{Reflection in mirrors}

\citet{gjurchinovski2004} discusses reflection of light from a
uniformly moving mirror and here I show how the {\tt lorentz} package
can illustrate some of his insights.  We are going to take the five
photons defined above and reflect them in an oblique mirror which is
itself moving at half the speed of light along the $x$-axis.  The
first step is to define the mirror {\tt m}, and the boost {\tt B}
corresponding to its velocity:
                                                              
\begin{Schunk}
\begin{Sinput}
> m <- c(1,1,1)
> B <- boost(as.3vel(c(0.5,0,0)))
\end{Sinput}
\end{Schunk}

Above, the three-vector $m$ is parallel to the normal vector of the
mirror and $B$ shows the Lorentz boost needed to bring it to rest.  We
are going to reflect these photons in this mirror.  The R idiom for
the reflection is performed using a sequence of transforms.  First,
transform the photons' four-momentum to a frame in which the mirror is
at rest:

\begin{Schunk}
\begin{Sinput}
> A
\end{Sinput}
\begin{Soutput}
     E   p_x    p_y    p_z
[1,] 1 0.990 0.0275 0.1375
[2,] 1 0.992 0.0551 0.1103
[3,] 1 0.993 0.0828 0.0828
[4,] 1 0.992 0.1103 0.0551
[5,] 1 0.990 0.1375 0.0275
\end{Soutput}
\begin{Sinput}
> (A <- as.4mom(A %*% t(B)))
\end{Sinput}
\begin{Soutput}
         E   p_x    p_y    p_z
[1,] 0.583 0.566 0.0275 0.1375
[2,] 0.582 0.569 0.0551 0.1103
[3,] 0.581 0.569 0.0828 0.0828
[4,] 0.582 0.569 0.1103 0.0551
[5,] 0.583 0.566 0.1375 0.0275
\end{Soutput}
\end{Schunk}

Above, see how the photons have lost energy because of a redshift (the
{\tt as.4mom()} function has no effect other than changing the column
names).  Next, reflect the photons in the mirror (which is at rest):

\begin{Schunk}
\begin{Sinput}
> (A <- reflect(A,m))
\end{Sinput}
\begin{Soutput}
         E    p_x    p_y    p_z
[1,] 0.583 0.0786 -0.460 -0.350
[2,] 0.582 0.0793 -0.434 -0.379
[3,] 0.581 0.0795 -0.407 -0.407
[4,] 0.582 0.0793 -0.379 -0.434
[5,] 0.583 0.0786 -0.350 -0.460
\end{Soutput}
\end{Schunk}

Above, see how the reflected photons have a reduced the x-component of
momentum; but have acquired a substantial $y$- and $z$- component.
Finally, we transform back to the original reference frame.  Observe
that this requires an {\em active} transform which means that we need
to use the matrix inverse of $B$:

\begin{Schunk}
\begin{Sinput}
> (A <- as.4mom(A %*% solve(t(B))))
\end{Sinput}
\begin{Soutput}
         E   p_x    p_y    p_z
[1,] 0.719 0.427 -0.460 -0.350
[2,] 0.718 0.427 -0.434 -0.379
[3,] 0.717 0.427 -0.407 -0.407
[4,] 0.718 0.427 -0.379 -0.434
[5,] 0.719 0.427 -0.350 -0.460
\end{Soutput}
\end{Schunk}

Thus in the original frame, the photons have lost about a quarter of
their energy as a result of a Doppler effect: the mirror was receding
from the source.  The photons have imparted energy to the mirror as a
result of mechanical work.  It is possible to carry out the same
operations in one line:

\begin{Schunk}
\begin{Sinput}
> A <- as.photon(as.3vel(cbind(0.9,1:5/40,5:1/40)))
> A %>% tcrossprod(B) %>% reflect(m) %>% tcrossprod(solve(B)) %>% as.4mom
\end{Sinput}
\begin{Soutput}
         E   p_x    p_y    p_z
[1,] 0.719 0.427 -0.460 -0.350
[2,] 0.718 0.427 -0.434 -0.379
[3,] 0.717 0.427 -0.407 -0.407
[4,] 0.718 0.427 -0.379 -0.434
[5,] 0.719 0.427 -0.350 -0.460
\end{Soutput}
\end{Schunk}

\subsection{Disco ball}

It is easy to define a disco ball, which is a sphere covered in
mirrors.  For the purposes of exposition, we will use a rather shabby
ball with only 7 mirrors:

\begin{Schunk}
\begin{Sinput}
> dfun <- function(n){matrix(rnorm(n*3),ncol=3) %>% sweep(1, sqrt(rowSums(.^2)),`/`)}
> (disco <- dfun(7))
\end{Sinput}
\begin{Soutput}
        [,1]    [,2]    [,3]
[1,] -0.8255  0.5638 -0.0246
[2,]  0.4893 -0.0811  0.8683
[3,] -0.0654  0.4519  0.8897
[4,] -0.6690 -0.2310 -0.7065
[5,] -0.7243 -0.6781 -0.1248
[6,] -0.5078 -0.4407  0.7402
[7,]  0.5224 -0.4929  0.6958
\end{Soutput}
\end{Schunk}

Then we define a unit-energy photon moving parallel to the x-axis, and
reflect it in the disco ball:

\begin{Schunk}
\begin{Sinput}
> p <- as.photon(c(1,0,0))
> reflect(p,disco)
\end{Sinput}
\begin{Soutput}
  E     p_x     p_y     p_z
x 1 -0.3630  0.9309 -0.0406
x 1  0.5212  0.0794 -0.8497
x 1  0.9914  0.0591  0.1164
x 1  0.1049 -0.3091 -0.9452
x 1 -0.0491 -0.9823 -0.1807
x 1  0.4843 -0.4476  0.7518
x 1  0.4542  0.5150 -0.7270
\end{Soutput}
\end{Schunk}

(above, {\tt p} is a photon moving along the x-axis; standard R
recycling rules imply that we effectively have one photon per mirror
in {\tt disco}.  See how the photons' energy is unchanged by the
reflection).  We might ask what percentage of photons are reflected
towards the source; but to do this we would need a somewhat more
stylish disco ball, here with 1000 mirrors:

\begin{Schunk}
\begin{Sinput}
> table(reflect(p,dfun(1000))[,2]>0) # should be TRUE with probability sqrt(0.5)
\end{Sinput}
\begin{Soutput}
FALSE  TRUE 
  294   706 
\end{Soutput}
\end{Schunk}

(compare the expected value of $1000/\sqrt{2}\simeq 707$).  But it is
perhaps more fun to consider a relativistic disco in which the mirror
ball moves at 0.5c:

\begin{Schunk}
\begin{Sinput}
> B <- boost(as.3vel(c(0.5,0,0)))
> p %>% tcrossprod(B) %>% reflect(disco) %>% tcrossprod(solve(B))
\end{Sinput}
\begin{Soutput}
      t      x       y       z
x 0.546 0.0913  0.5375 -0.0235
x 0.840 0.6808  0.0458 -0.4906
x 0.997 0.9943  0.0341  0.0672
x 0.702 0.4033 -0.1785 -0.5457
x 0.650 0.3006 -0.5671 -0.1043
x 0.828 0.6562 -0.2584  0.4340
x 0.818 0.6361  0.2973 -0.4197
\end{Soutput}
\end{Schunk}

Above, note the high energy of the photon in the third row.  This is
because the third mirror of {\tt disco} is such that the photon hits
it with grazing incidence; this means that the receding of the mirror
is almost immaterial.  Note further that a spinning disco ball would
give the same (instantaneous) results.

\subsection{Mirrors and rotation-boost coupling}

Consider the following situation: we take a bunch of photons which in
a certain reference frame are all moving (almost) parallel to the
$x$-axis.  Then we reflect the photons from a mirror which is moving
with a composition of pure boosts, and examine the reflected light in
their original reference frame.  The R idiom for this would be:

\begin{Schunk}
\begin{Sinput}
> sol(1)
\end{Sinput}
\begin{Soutput}
[1] 1
\end{Soutput}
\begin{Sinput}
> light_start <- as.photon(as.3vel(cbind(0.9,1:5/40,5:1/40)))
> m <- c(1,0,0)     # mirror normal to x-axis
> B1 <- boost(as.3vel(c(-0.5, 0.1, 0.0)))
> B2 <- boost(as.3vel(c( 0.2, 0.0, 0.0)))
> B3 <- boost(as.3vel(c( 0.0, 0.0, 0.6)))
> B <- B1 %*% B2 %*% B3   # matrix multiplication is associative!
> light <- light_start %*% t(B)
> light <- reflect(light,m)
> light <- as.4mom(light %*% solve(t(B)))
> light
\end{Sinput}
\begin{Soutput}
        E   p_x   p_y   p_z
[1,] 2.30 -2.11 0.119 0.920
[2,] 2.31 -2.13 0.147 0.897
[3,] 2.32 -2.14 0.175 0.874
[4,] 2.32 -2.15 0.203 0.849
[5,] 2.33 -2.16 0.230 0.824
\end{Soutput}
\end{Schunk}

See how the photons have picked up momentum in the $y$- and $z$-
direction, even though the mirror is oriented perpendicular to the
$x$-axis (in its own frame).  Again it is arguably preferable to use
pipes:

\begin{Schunk}
\begin{Sinput}
> light_start %>% tcrossprod(B) %>% reflect(m) %>% tcrossprod(solve(B)) %>% as.4mom
\end{Sinput}
\begin{Soutput}
        E   p_x   p_y   p_z
[1,] 2.30 -2.11 0.119 0.920
[2,] 2.31 -2.13 0.147 0.897
[3,] 2.32 -2.14 0.175 0.874
[4,] 2.32 -2.15 0.203 0.849
[5,] 2.33 -2.16 0.230 0.824
\end{Soutput}
\end{Schunk}

Compare when the speed of light is infinite:
\begin{Schunk}
\begin{Sinput}
> sol(Inf)
\end{Sinput}
\begin{Soutput}
[1] Inf
\end{Soutput}
\begin{Sinput}
> light_start <- as.photon(as.3vel(cbind(0.9,1:5/40,5:1/40)))
> B1 <- boost(as.3vel(c(-0.5, 0.1, 0.0)))
> B2 <- boost(as.3vel(c( 0.2, 0.0, 0.0)))
> B3 <- boost(as.3vel(c( 0.0, 0.0, 0.6)))
> B <- B1 %*% B2 %*% B3
> light_start
\end{Sinput}
\begin{Soutput}
     E   p_x    p_y    p_z
[1,] 0 0.990 0.0275 0.1375
[2,] 0 0.992 0.0551 0.1103
[3,] 0 0.993 0.0828 0.0828
[4,] 0 0.992 0.1103 0.0551
[5,] 0 0.990 0.1375 0.0275
\end{Soutput}
\begin{Sinput}
> light_start %>% tcrossprod(B) %>% reflect(m) %>% tcrossprod(solve(B)) %>% as.4mom
\end{Sinput}
\begin{Soutput}
     E    p_x    p_y    p_z
[1,] 0 -0.990 0.0275 0.1375
[2,] 0 -0.992 0.0551 0.1103
[3,] 0 -0.993 0.0828 0.0828
[4,] 0 -0.992 0.1103 0.0551
[5,] 0 -0.990 0.1375 0.0275
\end{Soutput}
\end{Schunk}

Note that, in the infinite light speed case, the energy of the photons
is zero (photons have zero rest mass); further observe that in this
classical case, the effect of the mirror is to multiply the $x$-momentum
by $-1$ and leave the other components unchanged, as one might expect
from a mirror perpendicular to $(1,0,0)$.

\section{Three-velocities}

In contrast to four-velocities, three-velocities do not form a group
under composition as the velocity addition law is not
associative~\citep{ungar2006}.  Instead, three-velocity composition
has an algebraic structure known as a {\em gyrogroup} (this
observation was the original motivation for the package).
\citeauthor{ungar2006} shows that the velocity addition law for
three-velocities is

\begin{equation}
\bu\oplus\bv=
\frac{1}{1+\bu\cdot\bv}
\left\{
\bu + \frac{\bv}{\gamma_\bu} + \frac{\gamma_\bu
\left(\bu\cdot\bv\right)\bu}{1+\gamma_\bu}
\right\}
\end{equation}
   
where~$\gamma_\bu=\left(1-\bu\cdot\bu\right)^{-1/2}$ and we are
assuming~$c=1$.  \citeauthor{ungar2006} goes on to show that, in
general, $\bu\oplus\bv\neq\bv\oplus\bu$
and~$(\bu\oplus\bv)\oplus\bw\neq\bu\oplus(\bv\oplus\bw)$.  He also
defines the binary operator~$\ominus$
as~$\bu\ominus\bv=\bu\oplus\left(-\bv\right)$, and implicitly
defines~$\ominus\bu\oplus\bv$ to be~$\left(-\bu\right)\oplus\bv$.  If
we have

\begin{equation}
\gyr{u}{v}\bx=-\left(\bu\oplus\bv\right)\oplus\left(\bu\oplus\left(\bv\oplus\bx\right)\right)
\end{equation}

then

\begin{eqnarray}
\bu\oplus\bv &=& \gyr{u}{v}\left(\bv\oplus\bu\right)\label{noncom}\\
\gyr{u}{v}\bx\cdot\gyr{u}{v}\by &=& \bx\cdot\by\label{doteq}\\
\gyr{u}{v}\left(\bx\oplus\by\right) &=& \gyr{u}{v}\bx\oplus\gyr{u}{v}\by\\
\left(\gyr{u}{v}\right)^{-1} &=& \left(\gyr{v}{u}\right)\label{gyrinv}\\
\bu\oplus\left(\bv\oplus\bw\right) &=&\left(\bu\oplus\bv\right)\oplus\gyr{u}{v}\bw\label{nonass1}\\
\left(\bu\oplus\bv\right)\oplus\bw &=&\bu\oplus\left(\bv\oplus\gyr{v}{u}\bw\right)\label{nonass2}
\end{eqnarray}

Consider the following R session:

\begin{Schunk}
\begin{Sinput}
> sol(1)
\end{Sinput}
\begin{Soutput}
[1] 1
\end{Soutput}
\begin{Sinput}
> u <- as.3vel(c(-0.7,+0.2,-0.3))
> v <- as.3vel(c(+0.3,+0.3,+0.4))
> w <- as.3vel(c(+0.1,+0.3,+0.8))
> x <- as.3vel(c(-0.2,-0.1,-0.9))
> u
\end{Sinput}
\begin{Soutput}
A vector of three-velocities (speed of light = 1)
        x   y    z
[1,] -0.7 0.2 -0.3
\end{Soutput}
\end{Schunk}

Here we have three-vectors {\tt u} etc.  We can see that {\tt u} and
{\tt v} do not commute:

\begin{Schunk}
\begin{Sinput}
> u+v
\end{Sinput}
\begin{Soutput}
A vector of three-velocities (speed of light = 1)
          x     y        z
[1,] -0.545 0.482 -0.00454
\end{Soutput}
\begin{Sinput}
> v+u
\end{Sinput}
\begin{Soutput}
A vector of three-velocities (speed of light = 1)
          x     y     z
[1,] -0.429 0.572 0.132
\end{Soutput}
\end{Schunk}

(the results differ).  We can use equation~\ref{noncom}
\begin{Schunk}
\begin{Sinput}
> (u+v)-gyr(u,v,v+u)
\end{Sinput}
\begin{Soutput}
A vector of three-velocities (speed of light = 1)
            x         y        z
[1,] 1.77e-16 -7.08e-16 1.23e-16
\end{Soutput}
\end{Schunk}

showing agreement to within numerical error.  It is also possible to
use the functional idiom in which we define {\tt f()} to be the
map~$\bx\mapsto\gyr{u}{v}\bx$.  In R:

\begin{Schunk}
\begin{Sinput}
> f <- gyrfun(u,v)
> (u+v)-f(v+u)    # should be zero
\end{Sinput}
\begin{Soutput}
A vector of three-velocities (speed of light = 1)
            x         y        z
[1,] 1.77e-16 -7.08e-16 1.23e-16
\end{Soutput}
\end{Schunk}

Function {\tt gyrfun()} is vectorized, which means that it plays
nicely with (R) vectors.  Consider

\begin{Schunk}
\begin{Sinput}
> u9 <- r3vel(9)
> u9
\end{Sinput}
\begin{Soutput}
A vector of three-velocities (speed of light = 1)
            x       y       z
 [1,] -0.6635 -0.1850 -0.0324
 [2,]  0.7232  0.3007  0.0448
 [3,] -0.4816 -0.2698  0.0903
 [4,]  0.7186  0.5215  0.1149
 [5,] -0.4988 -0.5469  0.4781
 [6,] -0.0446  0.7934 -0.4839
 [7,]  0.4225  0.6831 -0.2474
 [8,] -0.5546 -0.1326  0.4232
 [9,]  0.0366  0.0748  0.4407
\end{Soutput}
\end{Schunk}

Then we can create a vectorized gyrofunction:

\begin{Schunk}
\begin{Sinput}
> f <- gyrfun(u9,v)
> f(x)
\end{Sinput}
\begin{Soutput}
A vector of three-velocities (speed of light = 1)
            x         y      z
 [1,] -0.0282 -7.04e-02 -0.924
 [2,] -0.3568 -1.36e-01 -0.845
 [3,] -0.0692 -2.67e-02 -0.924
 [4,] -0.3503 -1.89e-01 -0.838
 [5,]  0.0444  1.59e-01 -0.913
 [6,] -0.2297 -4.52e-01 -0.776
 [7,] -0.3268 -3.10e-01 -0.811
 [8,]  0.0130  3.85e-06 -0.927
 [9,] -0.1409 -5.01e-02 -0.915
\end{Soutput}
\end{Schunk}

Note that the package vectorization is transparent when using syntactic sugar:

\begin{Schunk}
\begin{Sinput}
> u9+x
\end{Sinput}
\begin{Soutput}
A vector of three-velocities (speed of light = 1)
            x       y       z
 [1,] -0.7436 -0.2345 -0.5826
 [2,]  0.6411  0.2532 -0.6615
 [3,] -0.6319 -0.3444 -0.6273
 [4,]  0.6860  0.5266 -0.4421
 [5,] -0.6903 -0.6791 -0.0511
 [6,] -0.0952  0.7096 -0.6909
 [7,]  0.3115  0.6168 -0.6976
 [8,] -0.8231 -0.2462 -0.3690
 [9,] -0.2550 -0.0524 -0.7806
\end{Soutput}
\end{Schunk}

(here, the addition operates using R's standard recycling rules).

\subsection{Associativity}

Three velocity addition is not associative:

\begin{Schunk}
\begin{Sinput}
> (u+v)+w
\end{Sinput}
\begin{Soutput}
A vector of three-velocities (speed of light = 1)
          x     y     z
[1,] -0.465 0.655 0.501
\end{Soutput}
\begin{Sinput}
> u+(v+w)
\end{Sinput}
\begin{Soutput}
A vector of three-velocities (speed of light = 1)
          x     y     z
[1,] -0.549 0.667 0.416
\end{Soutput}
\end{Schunk}

But we can use equations~\ref{nonass1} and~\ref{nonass2}:

\begin{Schunk}
\begin{Sinput}
> (u+(v+w)) - ((u+v)+gyr(u,v,w))
\end{Sinput}
\begin{Soutput}
A vector of three-velocities (speed of light = 1)
            x         y         z
[1,] 6.92e-16 -1.38e-15 -6.92e-16
\end{Soutput}
\begin{Sinput}
> ((u+v)+w) - (u+(v+gyr(v,u,w)))
\end{Sinput}
\begin{Soutput}
A vector of three-velocities (speed of light = 1)
     x y        z
[1,] 0 0 5.35e-16
\end{Soutput}
\end{Schunk}

\subsection{Visualization of noncommutativity and nonassociativity of three-velocities}

Consider the following three-velocities:

\begin{Schunk}
\begin{Sinput}
> u <- as.3vel(c(0.4,0,0))
> v <- seq(as.3vel(c(0.4,-0.2,0)), as.3vel(c(-0.3,0.9,0)),len=20)
> w <- as.3vel(c(0.8,-0.4,0))
\end{Sinput}
\end{Schunk}

\begin{figure}[htbp]
  \begin{center}
\begin{Schunk}
\begin{Sinput}
> comm_fail1(u=u, v=v)
\end{Sinput}
\end{Schunk}
\includegraphics{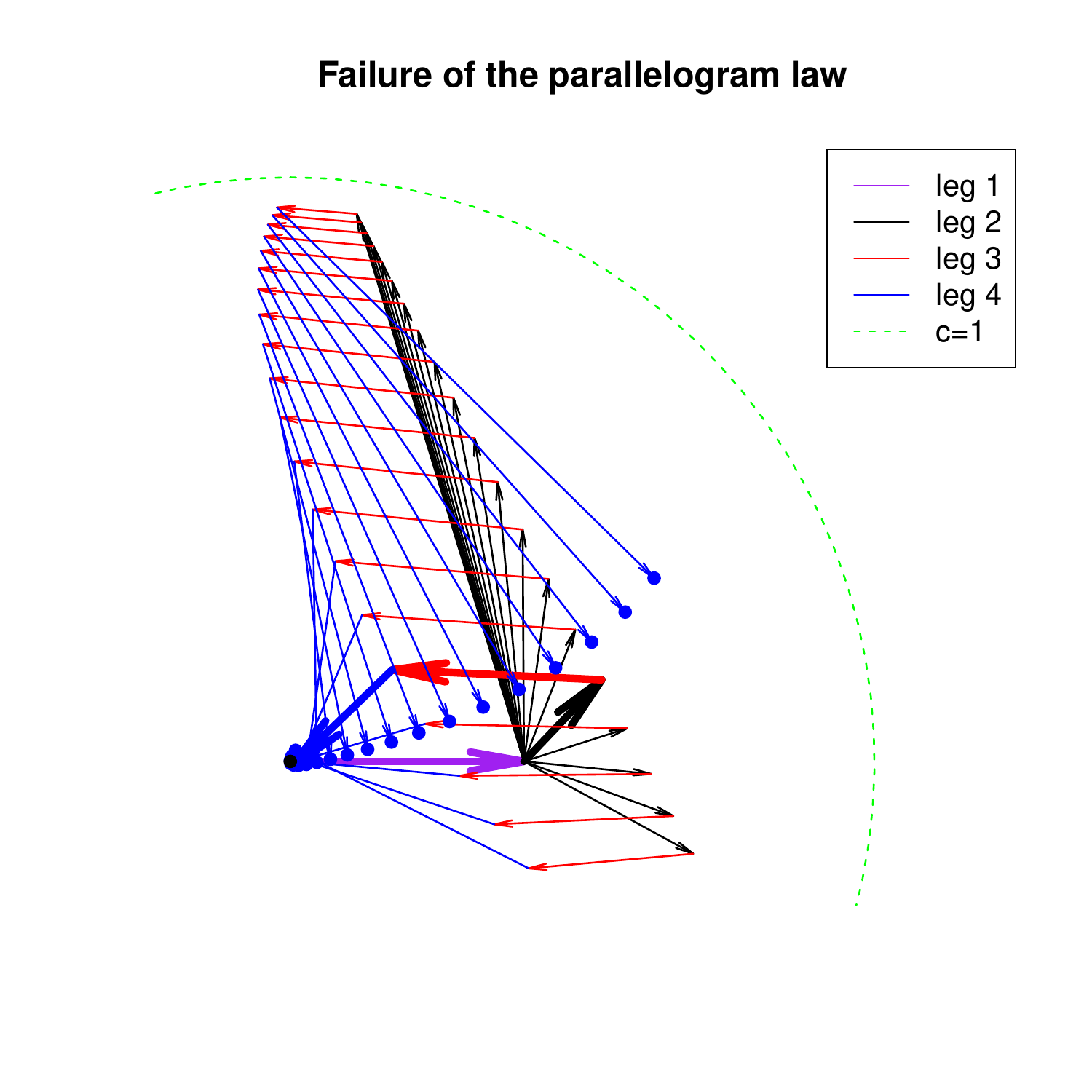}
\caption{Failure\label{comfail1} of the commutative law for velocity
  composition in special relativity.  The arrows show successive
  velocity boosts of $+\bu$ (purple), $+\bv$ (black), $-\bu$ (red),
  and~$-\bv$ (blue) for $\bu,\bv$ as defined above.  Velocity $\bu$ is
  constant, while $\bv$ takes a sequence of values.  If velocity
  addition is commutative, the four boosts form a closed
  quadrilateral; the thick arrows show a case where the boosts almost
  close and the boosts nearly form a parallelogram.  The blue dots
  show the final velocity after four successive boosts; the distance
  of the blue dot from the origin measures the combined velocity,
  equal to zero in the classical limit of low speeds.  The discrepancy
  becomes larger and larger for the faster elements of the sequence
  $\bv$}
  \end{center}
\end{figure}

\begin{figure}[htbp]
  \begin{center}
\begin{Schunk}
\begin{Sinput}
> comm_fail2(u=u, v=v)
\end{Sinput}
\end{Schunk}
\includegraphics{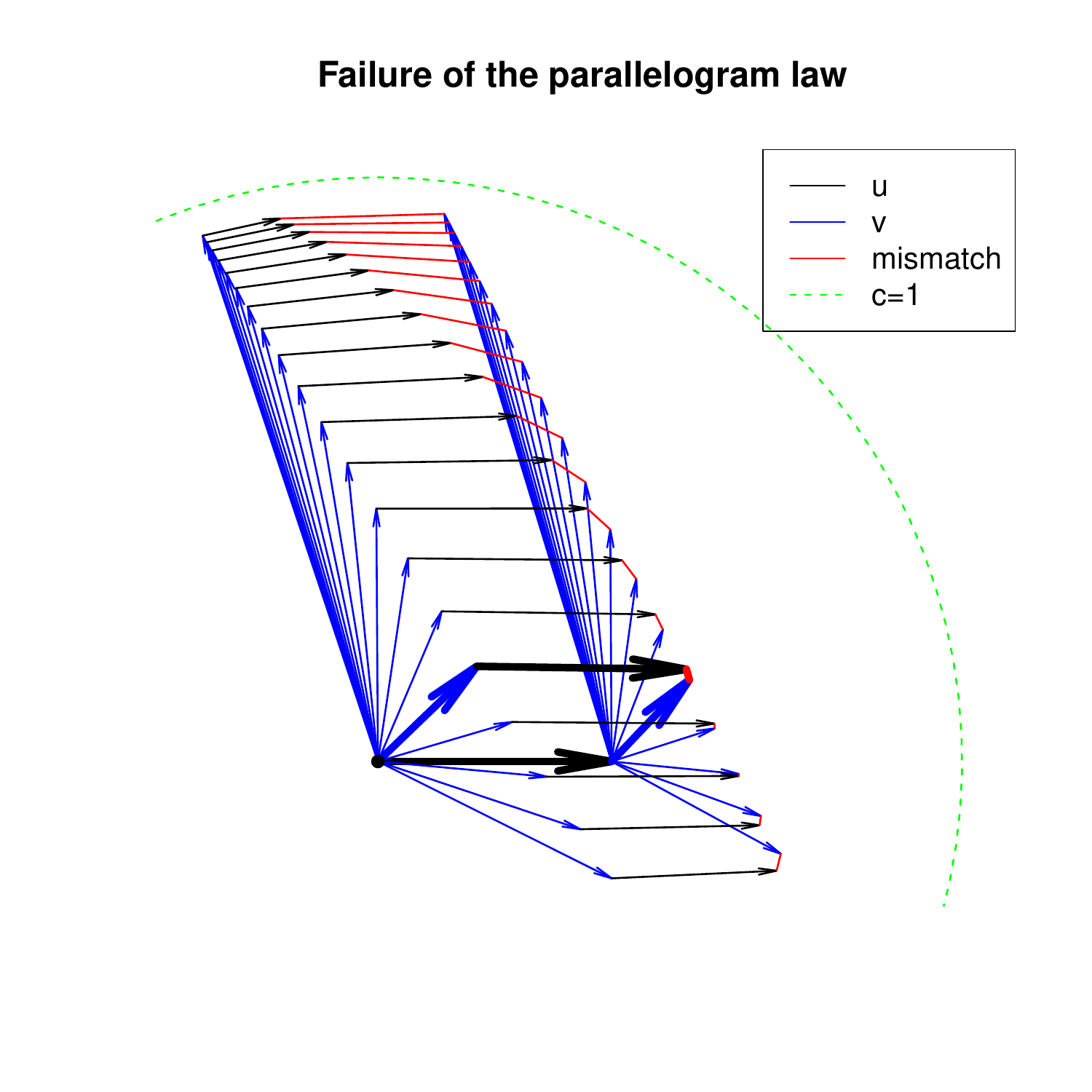}

\caption{Another view of the failure of the commutative
  law\label{comfail2} in special relativity.  The black arrows show
  velocity boosts of $\bu$ and the blue arrows show velocity boosts of
  $\bv$, with $\bu,\bv$ as defined above; $\bu$ is constant while
  $\bv$ takes a sequence of values.  If velocity addition is
  commutative, then $\bu+\bv=\bv+\bu$ and the two paths end at the
  same point: the parallelogram is closed.  The red lines show the
  difference between $\bu+\bv$ and $\bv+\bu$}
  \end{center}
\end{figure}

\begin{figure}[htbp]
  \begin{center}
\begin{Schunk}
\begin{Sinput}
> ass_fail(u=u, v=v, w=w, bold=10)
\end{Sinput}
\end{Schunk}
\includegraphics{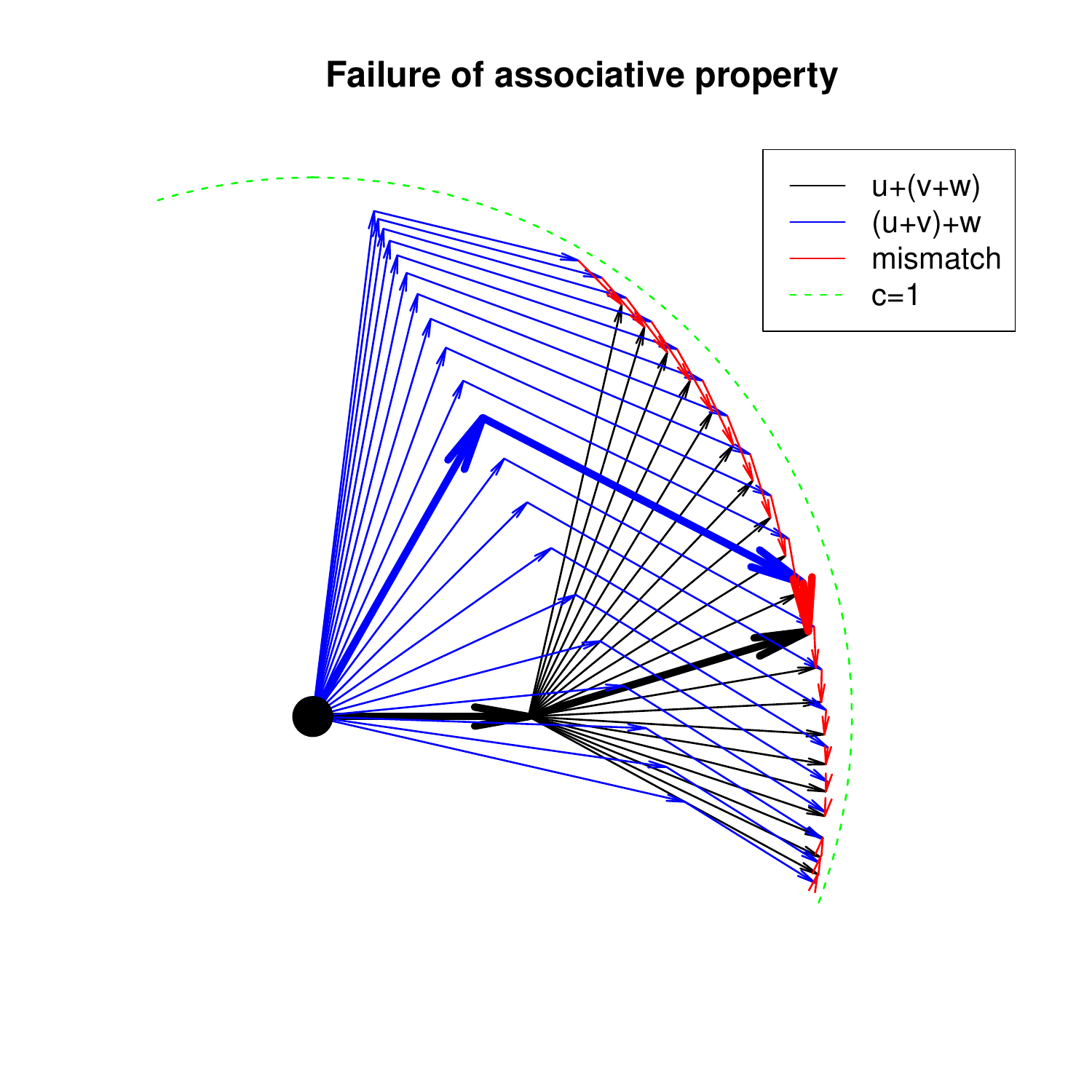}
\caption{Failure of the associative law \label{assfail} for velocity
  composition in special relativity.  The arrows show successive
  boosts of $\bu$ followed by $\bv+\bw$ (black lines), and $\bu+\bv$
  followed by $\bw$ (blue lines), for $\bu$, $\bv$, $\bw$ as defined
  above; $\bu$ and $\bw$ are constant while $\bv$ takes a sequence of
  values. The mismatch between $\bu+\left(\bv+\bw\right)$ and
  $\left(\bu+\bv\right)+\bw$ is shown in red}
  \end{center}
\end{figure}

Objects~$\bv$ and $\bw$ are single three-velocities, and object $\bv$
is a vector of three velocities.  We can see the noncommutativity of
three velocity addition in figures~\ref{comfail1} and~\ref{comfail2},
and the nonassociativity in figure~\ref{assfail}.

\subsection[The magrittr package: pipes]{The {\tt magrittr} package: pipes}

Three velocities in the {\tt lorentz} package work nicely with
{\tt magrittr}.  If we define

\begin{Schunk}
\begin{Sinput}
>  u <- as.3vel(c(+0.5,0.1,-0.2))
>  v <- as.3vel(c(+0.4,0.3,-0.2))
>  w <- as.3vel(c(-0.3,0.2,+0.2))
\end{Sinput}
\end{Schunk}

Then pipe notation operates as expected:

\begin{Schunk}
\begin{Sinput}
> jj1 <- u %>% add(v)
> jj2 <- u+v
> speed(jj1-jj2)
\end{Sinput}
\begin{Soutput}
[1] 2.21e-16
\end{Soutput}
\end{Schunk}

The pipe operator is left associative:

\begin{Schunk}
\begin{Sinput}
> jj1 <- u %>% add(v) %>% add(w)
> jj2 <- (u+v)+w
> speed(jj1-jj2)
\end{Sinput}
\begin{Soutput}
[1] 7.39e-17
\end{Soutput}
\end{Schunk}

If we want right associative addition, the pipe operator needs
brackets:

\begin{Schunk}
\begin{Sinput}
> jj1 <- u %>% add(v %>% add(w))
> jj2 <- u+(v+w)
> speed(jj1-jj2)
\end{Sinput}
\begin{Soutput}
[1] 3.45e-17
\end{Soutput}
\end{Schunk}

\subsection{Numerical verification}

Here I provide numerical verification of equations~\ref{noncom}
to~\ref{nonass2}.  If we have

\begin{Schunk}
\begin{Sinput}
> x <- as.3vel(c(0.7, 0.0, -0.7))
> y <- as.3vel(c(0.1, 0.3, -0.6))
> u <- as.3vel(c(0.0, 0.8, +0.1))   # x,y,u: single three-velocities
> v <- r3vel(5,0.9)
> w <- r3vel(5,0.8)                 # v,w: vector of three-velocities
> f <- gyrfun(u,v)
> g <- gyrfun(v,u)
\end{Sinput}
\end{Schunk}

Then we can calculate the difference between the left hand side and
right hand side numerically:
  
\begin{Schunk}
\begin{Sinput}
> max(speed((u+v) - f(v+u)))              # equation 3
\end{Sinput}
\begin{Soutput}
[1] 3.49e-13
\end{Soutput}
\begin{Sinput}
> max(abs(prod3(f(x),f(y)) - prod3(x,y))) # equation 4
\end{Sinput}
\begin{Soutput}
[1] 5.94e-15
\end{Soutput}
\begin{Sinput}
> max(speed(f(x+y) - (f(x)+f(y))))        # equation 5
\end{Sinput}
\begin{Soutput}
[1] 3.81e-12
\end{Soutput}
\begin{Sinput}
> max(speed(f(g(x)) - g(f(x))))           # equation 6
\end{Sinput}
\begin{Soutput}
[1] 7.3e-13
\end{Soutput}
\begin{Sinput}
> max(speed((u+(v+w)) - ((u+v)+f(w))))    # equation 7
\end{Sinput}
\begin{Soutput}
[1] 6.25e-14
\end{Soutput}
\begin{Sinput}
> max(speed(((u+v)+w) - (u+(v+g(w)))))    # equation 8
\end{Sinput}
\begin{Soutput}
[1] 2.84e-14
\end{Soutput}
\end{Schunk}

(all zero to numerical precision).

\section{Conclusions}

The {\tt lorentz} package furnishes some functionality for
manipulating four-vectors and three-velocities in the context of
special relativity.  The R idiom is relatively natural and the package
has been used to illustrate different features of relativistic
kinematics.  The package leverages the powerful R programming language
to conduct a systematic search for a gyrodistributive law, without
success.  If such a law exists, it is complicated: it is not one of
the 688128 natural forms considered in the search.

\bibliographystyle{apalike}
\bibliography{lorentz_arxiv}

\appendix
\section{A systematic search for a distributive law for three-velocities}

Ungar~\cite{ungar1997} states:

\begin{quote}
It is hoped that one day a gyrodistributive law connecting [three
velocity addition] $\oplus$ and [scalar multiplication] $\odot$ will
be discovered.  If [it] exists, it is expected to be the standard
distributive law relaxed by Thomas gyration in some unexpected way.
\end{quote}

Here, I make a systematic attempt to find a distributive law for
three-velocities, that is, finding an expression for
$r\odot(\bu\oplus\bv)$ where $\bu$ and $\bv$ are three-velocities and
$r$ is a real number.  We define a large number of possible
combinations of $\bu$, $\bv$, and $r$ using {\tt lorentz} package
idiom.  These are calculated and the results compared with the exact
value.  If there is an exact expression, this will have zero
discrepancy.

I have tried to enumerate the various combinations of $\bu,\bv,r$.  Signs
are permuted in {\tt every\_sign()} and {\tt all3()} and {\tt
  all3brack()}.  If one of the many combinations (the current total
stands at 688128) represents the RHS of a putative distributive law,
the corresponding element of variable {\tt badness}, defined at the
end of this script, will be zero.

Function {\tt possible()}, defined below, returns a long vector of
possible expressions that might form the RHS for a distributive law.
The basic idea is that $r\odot(\bu\oplus\bv) = r\odot\bu+r\odot\bv + X$
(where $X$ is a correction term) but because of noncommutativity and
nonassociativity, this gets complicated.  Function {\tt possible()}
returns a vector that includes expressions such as

\begin{Schunk}
\begin{Sinput}
## ru  + (rv  - r*gyr[ru, v,u+v])
## ru  + (rv  - r*gyr[ru, v,v+u])
## rv  + (ru  - r*gyr[su, v,u+v])
## rv  + (ru  - r*gyr[su, v,u+v])
## gyr[ru,-v,v+u] + (ru+rv)
## (gyr[ru,-v,v+u] + ru) +rv
## s*gyr[ru,-v,v] + (ru+rv)
## (s*gyr[ru,-v,v] + ru)  + rv
## ru + (s*gyr[ru,u,v] + rv)
## (s*gyr[ru,-v,v] + ru)  + rv
\end{Sinput}
\end{Schunk}

etc etc etc.  Here, $r$ is a scalar, $s=1/r$, and $u,v$ are
three-velocities; {\tt ru} and {\tt su} represent $r\odot\bu$ and
$s\odot\bu$.  Note the variety of orders (three-velocity addition is
not commutative), different bracketing (three-velocity addition is not
associative), and use of $r$ or $1/r$ in different places in the
formula.  The {\tt partitions} package~\cite{hankin2006} is used to
ensure that the scheme may easily be improved in the future.

\begin{Schunk}
\begin{Sinput}
> library("partitions") # needed for perms()
> u <- r3vel(1,0.4)
> v <- r3vel(1,0.5)
> r <- 2
> `possible` <- function(u,v,r){ 
+  # In function possible(), u,v are three-velocities and r is a real
+  # number.  Function possible() returns a vector of 64*10752=688128
+  # three-velocities [the '64' is from combinations of r, 1/r as in jj
+  # below; the 10752 is from function f(), defined inside possible()]
+ 
+   f <- function(r1,r2,r3){
+     ## In function f(), r1,r2,r3 are real numbers; function f() has 4*14
+     ## = 56 lines, so returns 56*192=10752 three-velocities [192 from
+     ## every_sign()]
+ 
+     ## NB: Inside f(), r,u,v come from possible()'s scope
+     c(
+         every_sign(r*u,r*v, r1*u     ,r2*v  , u+v,r3),
+         every_sign(r*u,r*v, r1*u     ,r2*v  , v+u,r3),
+         every_sign(r*u,r*v, r1*v     ,r2*u  , u+v,r3),
+         every_sign(r*u,r*v, r1*v     ,r2*u  , v+u,r3),
+ 
+         every_sign(r*u,r*v, r1*u     ,r2*v  , u-v,r3),
+         every_sign(r*u,r*v, r1*u     ,r2*v  , v-u,r3),
+         every_sign(r*u,r*v, r1*v     ,r2*u  , u-v,r3),
+         every_sign(r*u,r*v, r1*v     ,r2*u  , v-u,r3),
+         
+         every_sign(r*u,r*v, r1*u     ,r2*v+u, u+v,r3),
+         every_sign(r*u,r*v, r1*u     ,r2*v+u, v+u,r3),
+         every_sign(r*u,r*v, r1*u     ,r2*u+v, u+v,r3),
+         every_sign(r*u,r*v, r1*u     ,r2*u+v, v+u,r3),
+ 
+         every_sign(r*u,r*v, r1*u     ,r2*v-u, u+v,r3),
+         every_sign(r*u,r*v, r1*u     ,r2*v-u, v+u,r3),
+         every_sign(r*u,r*v, r1*u     ,r2*u-v, u+v,r3),
+         every_sign(r*u,r*v, r1*u     ,r2*u-v, v+u,r3),
+ 
+         every_sign(r*u,r*v, r1*u     ,r2*v+u, u-v,r3),
+         every_sign(r*u,r*v, r1*u     ,r2*v+u, v-u,r3),
+         every_sign(r*u,r*v, r1*u     ,r2*u+v, u-v,r3),
+         every_sign(r*u,r*v, r1*u     ,r2*u+v, v-u,r3),
+         
+         every_sign(r*u,r*v, r1*u     ,r2*v-u, u-v,r3),
+         every_sign(r*u,r*v, r1*u     ,r2*v-u, v-u,r3),
+         every_sign(r*u,r*v, r1*u     ,r2*u-v, u-v,r3),
+         every_sign(r*u,r*v, r1*u     ,r2*u-v, v-u,r3),
+ 
+         
+         every_sign(r*u,r*v, r1*u     ,v+r2*u, u+v,r3),
+         every_sign(r*u,r*v, r1*u     ,v+r2*u, v+u,r3),
+         every_sign(r*u,r*v, r1*u     ,u+r2*v, u+v,r3),
+         every_sign(r*u,r*v, r1*u     ,u+r2*v, v+u,r3),
+         
+         every_sign(r*u,r*v, r1*u     ,v+r2*u, u-v,r3),
+         every_sign(r*u,r*v, r1*u     ,v+r2*u, v-u,r3),
+         every_sign(r*u,r*v, r1*u     ,u+r2*v, u-v,r3),
+         every_sign(r*u,r*v, r1*u     ,u+r2*v, v-u,r3),
+         
+         every_sign(r*u,r*v, r1*u     ,v-r2*u, u+v,r3),
+         every_sign(r*u,r*v, r1*u     ,v-r2*u, v+u,r3),
+         every_sign(r*u,r*v, r1*u     ,u-r2*v, u+v,r3),
+         every_sign(r*u,r*v, r1*u     ,u-r2*v, v+u,r3),
+         
+         every_sign(r*u,r*v, r1*u     ,v-r2*u, u-v,r3),
+         every_sign(r*u,r*v, r1*u     ,v-r2*u, v-u,r3),
+         every_sign(r*u,r*v, r1*u     ,u-r2*v, u-v,r3),
+         every_sign(r*u,r*v, r1*u     ,u-r2*v, v-u,r3),
+         
+         every_sign(r*u,r*v, r1*u+r2*v,v     , u+v,r3),
+         every_sign(r*u,r*v, r1*u+r2*v,v     , v+u,r3),
+         every_sign(r*u,r*v, r1*v+r2*u,u     , u+v,r3),
+         every_sign(r*u,r*v, r1*v+r2*u,u     , v+u,r3),
+         
+         every_sign(r*u,r*v, r1*u+r2*v,v     , u-v,r3),
+         every_sign(r*u,r*v, r1*u+r2*v,v     , v-u,r3),
+         every_sign(r*u,r*v, r1*v+r2*u,u     , u-v,r3),
+         every_sign(r*u,r*v, r1*v+r2*u,u     , v-u,r3),
+         
+         every_sign(r*u,r*v, r1*u-r2*v,v     , u+v,r3),
+         every_sign(r*u,r*v, r1*u-r2*v,v     , v+u,r3),
+         every_sign(r*u,r*v, r1*v-r2*u,u     , u+v,r3),
+         every_sign(r*u,r*v, r1*v-r2*u,u     , v+u,r3),
+         
+         every_sign(r*u,r*v, r1*u-r2*v,v     , u-v,r3),
+         every_sign(r*u,r*v, r1*u-r2*v,v     , v-u,r3),
+         every_sign(r*u,r*v, r1*v-r2*u,u     , u-v,r3),
+         every_sign(r*u,r*v, r1*v-r2*u,u     , v-u,r3)
+ 
+     )
+   }   # function f() closes
+ 
+   r11r0 <- c(r,1,1/r,0)
+   jj <- as.matrix(expand.grid(r11r0,r11r0,r11r0))
+   ## jj has 4^3=64 rows
+   out <- u
+   for(i in seq_len(nrow(jj))){
+       ## cat(paste(i," / ",nrow(jj),"\n",sep="")) # for debugging
+     out <- c(out,f(jj[i,1],jj[i,2],jj[i,3]))
+   } 
+   return(out)  # out has 64*110752=688128 elements
+ } # function possible() closes
> `every_sign` <- function(a1,a2,a3,a4,a5,r){
+  # Function every_sign() has 16 lines; given 5 three-velocities and a
+  # scalar, function every_sign() returns a vector of 16*12=192
+  # three-velocities [the 12 is from all3()]
+   c(
+       all3(c(a1 , a2 , +r*gyr(+a3,+a4,+a5))),
+       all3(c(a1 , a2 , +r*gyr(+a3,+a4,-a5))),
+       all3(c(a1 , a2 , +r*gyr(+a3,-a4,+a5))),
+       all3(c(a1 , a2 , +r*gyr(+a3,-a4,-a5))),
+       all3(c(a1 , a2 , +r*gyr(-a3,+a4,+a5))),
+       all3(c(a1 , a2 , +r*gyr(-a3,+a4,-a5))),
+       all3(c(a1 , a2 , +r*gyr(-a3,-a4,+a5))),
+       all3(c(a1 , a2 , +r*gyr(-a3,-a4,-a5))),
+       all3(c(a1 , a2 , -r*gyr(+a3,+a4,+a5))),
+       all3(c(a1 , a2 , -r*gyr(+a3,+a4,-a5))),
+       all3(c(a1 , a2 , -r*gyr(+a3,-a4,+a5))),
+       all3(c(a1 , a2 , -r*gyr(+a3,-a4,-a5))),
+       all3(c(a1 , a2 , -r*gyr(-a3,+a4,+a5))),
+       all3(c(a1 , a2 , -r*gyr(-a3,+a4,-a5))),
+       all3(c(a1 , a2 , -r*gyr(-a3,-a4,+a5))),
+       all3(c(a1 , a2 , -r*gyr(-a3,-a4,-a5)))
+   )
+ }
> `all3brack` <- function(x){  # If [abc] = c(x[1],x[2],x[3]), function
+                              # all3brack() returns a+(b+c) and (a+b)+c
+   c(
+       x[1]+(x[2]+x[3])   ,
+       (x[1]+x[2])+x[3]
+   )
+ }
> `all4brack` <- function(x){
+   ## Returns the 5 different ways to bracket 4 objects.  Function
+   ## all4vbrack() not currently used in this script.
+   c(
+   (x[1]+x[2])+(x[3]+x[4])   ,
+   ((x[1]+x[2])+x[3])+x[4]   ,
+   (x[1]+(x[2]+x[3]))+x[4]   ,
+   x[1]+((x[2]+x[3])+x[4])   ,
+   x[1]+(x[2]+(x[3]+x[4]))
+   )
+ }
> `all4` <- function(x){  
+  ## Every possible way of combining 4 three-velocities.  Function
+  ## all4() is not currently used in this script
+   stopifnot(length(x)==4)
+   out <- threevel(0)
+   jj <- perms(4)
+   for(i in seq_len(ncol(jj))){
+     out <- c(out,all4brack(x[jj[,i]]))
+   }
+   return(out)
+ }
> `all3` <- function(x){   # every possible way of combining 3
+                          # three-velocities.  Function; if x=c(a,b,c)
+                          # with a,b,c three-velocities then all3()
+                          # returns a vector of length 2*3!=12
+ 
+   stopifnot(length(x)==3)
+   out <- threevel(0)
+   jj <- perms(3)
+   for(i in seq_len(ncol(jj))){
+     out <- c(out,all3brack(x[jj[,i]]))
+   }
+   return(out)
+ }
\end{Sinput}
\end{Schunk}

We can now run the diagnostic:

\begin{Schunk}
\begin{Sinput}
> badness <- prod3(r*(u+v) - possible(u,v,r))  # badness == 0 for perfect law
> print(min(badness))
\end{Sinput}
\begin{Soutput}
[1] 0.00216
\end{Soutput}
\end{Schunk}

We see that this systematic sweep through function space does not
include a distributive law; the minimum error (at about 0.002) is far
greater than one would expect for a correct law (if it exists) which
would be subject to numerical roundoff error only.

\end{document}